\documentstyle[12pt,mysty]{article}
\begin{document}
\newcommand{\abstrait}{We examine a model of hadronic diffractive scattering
which interpolates between perturbative QCD and non-perturbative fits.
We restrict the perturbative QCD resummation to the large
transverse momentum region, and use a simple Regge-pole parametrization
in the infrared region.
This picture allows us to account for existing data, and to estimate the size
of the perturbative contribution to future diffractive measurements.
At LHC and SSC energies, we find that a cut-off BFKL equation can lead to
a measurable perturbative component in traditionally soft processes. In
particular, we show that the total $pp$ cross section could become as large
as 228 mb (160 mb) and the $\rho$
parameter as large as 0.23 (0.24) at the SSC (LHC).}
\begin{titlepage}
\begin{flushright}
McGill/92--37\\
September 1992
\end{flushright}
\vspace{.5in}
\begin{center}
\renewcommand{\thefootnote}{*}
\large\bf{Can perturbative QCD predict a substantial part
of diffractive LHC/SSC physics?}

\vspace{.2in}
{J.R. Cudell\footnote{cudell@hep.physics.mcgill.ca} and B. Margolis}

\vspace{.2in}
{\it Physics Department, McGill University\\
Montr\'eal, Qu\'ebec H3A 2T8, Canada\rm}
\vspace{.7 in}

{\bf Abstract}
\end{center}
\abstrait
\end{titlepage}
\newpage
\section{Introduction}
\label{intro}

\noindent
As the energy of hadronic colliders increases, diffractive
scattering will play an increasingly
important role. On the discovery side, it will produce the highest mass states
accessible at future colliders\cite{Bialas}, and the physics of rapidity gaps
might make their detection feasible\cite{FAD}.
On the background side, most interesting events will emerge
from the small-$x$ region, and will contain an appreciable  ``minijet''
structure, so that soft parton scattering and evolution has to be modelled to
optimize the detection of new physics.

As lattice calculations can deal only with static problems, the
only fundamental tool
that we have so far to deal with QCD scattering is
perturbation theory, and resummation techniques\cite{BFKL,CollinsL} have
pushed the
perturbative limit to the small-x region. However, it seems that, at present
energies, these efforts have failed to reproduce soft data\cite{RossH}. This
failure has
been ascribed to the intrinsically non-perturbative nature of the problem, and
simple models have been proposed to extrapolate present measurements to higher
energies\cite{Donnachie,TTWu,Margolis}. However, the details of the process
cannot be predicted through this approach.

We thus want to address the following question: what fraction of events at
future colliders can be understood by present perturbative techniques?
The first step is to find a simple parametrization of the data, for which we
use one of the existing models. We assume that this describes the infrared
part of the QCD ladders, for which the gluon transverse momentum $k_T$ is
smaller than some cutoff $Q_0$. We then evolve this term via perturbative
resummation techniques\cite{BFKL},
using gluons with $k_T>Q_0$, so that we are sure that perturbative QCD
is valid.
This approach interpolates between the purely perturbative ladders ($Q_0=0$)
and the purely non-perturbative models ($Q_0=\infty$).

We limit ourselves to the most general features that one can expect
from such an
evolution, and do not attempt to make an explicit model. We simply assume that
the infrared region couples to the perturbative one through an unknown vertex.
For a given $Q_0$, present data constrain the size of this vertex
and one can predict an upper bound on the perturbative contribution to the
hadronic amplitude at higher energies.
As the QCD equations are simpler at zero momentum transfer,
we consider only the total
cross section and the ratio of real to imaginary part of the forward scattering
amplitude, the $\rho$ parameter. Even then, as the exchange will
involve at least two gluons, it is possible to demand that both have large
transverse momenta, which add up to zero. The perturbative evolution then can
lead to a ``gluon bomb" which remains dormant in the data up to present
energies, but which can bring large observable corrections at future
colliders.

In the next section, we give a simple model for soft physics at $t=0$, which we
call the soft pomeron. We then briefly outline the BFKL equation \cite{BFKL}
and mention its solutions, which are very far from reproducing the data. We
then show how one can make a very general model evolving soft physics to
higher values of $\log s$ and constrain it using existing data for
$\sigma_{tot}$ and $\rho$. We then show that soft physics at the SSC and the
LHC could have a substantial perturbative component.

\section{Data: the soft pomeron}

\noindent As explained above, we shall concentrate on the hadronic amplitude
 ${\cal A}\sl(s,t=0)$ describing the elastic scattering of
pp and p$\rm\bar p$ with center-of-mass energy $\sqrt{s}$ and squared
momentum transfer $t=0$. This amplitude is known experimentally: we normalize
it so that its imaginary part is $s$ times the total cross section; the ratio
of its real and imaginary parts is by definition $\rho$.

The most economical fit, inspired by Regge theory,
is a sum of two simple Regge poles:
\begin{equation}
 {{\cal A}(s,t)\over s}=(a\pm i b) s^{\epsilon_m+\alpha_m' t}+C_0
s^{\epsilon_0+\alpha' t}
\end{equation}
with $a$, $b$, $C_0$ constants independent of s.
The phase of the amplitude is obtained by the imposition of $s$ to $u$
crossing symmetry. The first term has a universal part ($a$) representing
$f$ and $a_2$ exchange, and a part ($b$) changing sign between p and $\rm\bar
p$
scattering, which comes from $\rho$ and $\omega$ exchange. The second term
($C_0$) is responsible for the rise in $\sigma_{tot}$ and is referred to as the
``soft pomeron''. Its only obvious problem is the eventual violation of the
Froissart bound. Therefore, we also consider a unitarized version, for which we
eikonalize the second term of Equation (1).

We give the best fit values of the parameters and the $\chi^2/$d.o.f. in
Table 1.
\begin{center}
\begin{tabular} {|l|l|l|l|l|}   \hline
parameter                       & pole fit      & eikonal fit \\ \hline
$a$                             &124 mb         & 141 mb       \\
$b$                             &32.5 mb        & 35.1 mb       \\
$\epsilon_m$                    &-0.474          & -0.469       \\
$C_0$                           &21.6 mb        & 24.0 mb     \\
$\epsilon_0$		        &0.085          & 0.093       \\
$\alpha'$                       &  -            & 0.251       \\ \hline
$\chi^2/d.o.f.$                 &1.03           & 1.08        \\ \hline
$\sigma_{tot}$ at the SSC (LHC) &125 (107) mb   & 117 (102) mb\\
$\rho$ at the SSC (LHC)         &0.131 (0.131)    & 0.116 (0.113) \\ \hline
\end{tabular}
\end{center}
\noindent Table 1: Values of the parameters of Equation (1) that result from a
least-$\chi^2$ fit to data at $t=0$.

The pole fit is shown by the lower curve of Figure 1 and the eikonalized one
by the lowest curve of Figure 2. Both fits reproduce the data \cite{data},
from $\sqrt{s}=$ 10 Gev to 1800 GeV
with $\chi^2/$d.o.f. very close to 1.  The only failure is
the UA4 value for $\rho$,
which is not reproduced by most models, and for which further
experimental confirmation seems to be
needed. It is a curious fact that the eikonalized fit chooses the conventional
value of the pomeron slope $\alpha'$ which is normally
derived from other constraints\cite{Donnachie}.
Also, notice that unitarization does not make a big difference, and that even
at the SSC, the difference between the two fits is only 8 mb.

Other parametrizations are possible, {\it e.g.} \cite{TTWu,Margolis},
and as shown by the proponents of this one \cite{Donnachie}, multiple Regge
exchanges are essential to describe the data at nonzero $t$. However, as
we limit ourselves here to the zero momentum transfer case for which
the corrections are small, and as this simple form is particularly
well suited for our purpose, we shall adopt it in the following as a
starting point for the QCD evolution.

\section{Theory: the hard pomeron}

\noindent In order to describe total cross sections within
the context of perturbative
QCD, one can try, for $s\rightarrow\infty$, to isolate the
leading contributions
and to resum them. This is made possible by the fact that perturbative QCD is
infrared finite in the leading $\log s$ approximation and in the
colour-singlet channel. This suggests that very small momenta might not
matter, and that one could use perturbation theory.

Such a program has been developed by BFKL \cite{BFKL}. In a nutshell,
one can show that, when considering gluon diagrams only,
the amplitude is a sum of terms $T_n$ of order $(\log s)^n$
and that terms of order $(\log s)^n$ are related to terms of order
$(\log s)^{n-1}$ by an
integral operator that does not depend on $n$, and that we shall write
$\hat K$:
\begin{eqnarray}
&&T_{n+1}(s,k_T^2)=\hat K T_n(s',k_T'^2)\nonumber\\
&=&{3\alpha_S\over \pi} k_T^2
\int_{s_0}^s {ds'\over s'}\int {dk_T'^2\over
k_T'^2}[{T_n(s',k_T'^2)-T_n(s',k_T^2)\over
|k_T^2-k_T'^2|}+{T_n(s',k_T^2)\over\sqrt{k_T^2+
4 k_T'^2}}].
\end{eqnarray}
This leads to:
\begin{equation}
T_{\infty}=\sum_n T_n=T_0+\hat K T_{\infty}
\end{equation}
This is the BFKL equation at $t=0$. Its extension to nonzero $t$ is known, but
too complicated to handle analytically. We limit ourselves here to the zero
momentum transfer case.

In this regime, the BFKL equation (3) possesses two classes of solutions.
First of all, at
fixed $\alpha_S$, the resummed amplitude is a Regge cut
instead of a simple pole:
$T_\infty\approx\int d\nu s^{N(\nu)}$, with a leading behaviour given by
\begin{equation}
N_{max}=1+{12 \log 2\over\pi}\alpha_S
\end{equation}
Even for a small $\alpha_S$, say of order of 0.2, this leads to a big intercept
$N_{max}\approx 1.5$. As this is much too big to accomodate the data, and as a
cut rather than a pole leads to problems with quark counting, subleading
terms were added via the running of the coupling constant. It was first
claimed that such terms would discretize the cut and turn it into a series of
poles \cite{LipatovKirchner}, but further work has shown that the cut structure
remains \cite{RossH,RossD}. However, the leading singularity
is slightly reduced, and
one can derive the bound \cite{CollinsK}
\begin{equation}
N_{max}>1+{3.6\over\pi}\alpha_S
\end{equation}
Again, for values of $\alpha_S$ of the order of 0.2, this leads to an
intercept of the order of 1.23.

So, we reach a contradiction: on the one hand, the data demands that the
amplitude rises more slowly than $s^{1+\epsilon_0}$, with
$\epsilon_0<0.1$; on the other hand,
perturbative resummation leads to a power $s^{1+\epsilon_p}$, with
$\epsilon_p>0.23$. The difference between
the two is a factor 3 in the total cross section at the Tevatron. The
resolution of this problem is far from clear, and one can envisage the
implementation of some non-perturbative effects within the BFKL equation
\cite{RossH}. Rather than trying to understand $\epsilon_0$,
we shall here take a
much simpler approach, {\it i.e.}
assume a low-$k_T$, low-$s$ behaviour consistent with the
data, and see what general features its perturbative evolution might exhibit.

The idea is to cut off Equation (3) by imposing $k_T^2>Q_0^2$, with $Q_0$
big enough for perturbation theory to apply, so that one uses the perturbative
resummation only at short distances. Furthermore, one takes $T_0\sim s^{1+
\epsilon_0}$ as the non-perturbative driving term, valid for $k_T^2<Q_0^2$.
This cut-off equation has
been recently solved by Collins and Landshoff \cite{CollinsL} in the case of
deep inelastic scattering.
Most of their results and approximations can be carried over
to the hadron-hadron scattering case,
and we shall give here the basic features of the solution in this case.

First of all, the hadronic amplitude can be thought of as the convolution of
two form factors times a resummed QCD gluonic amplitude obeying a cut-off BFKL
equation.
\begin{equation}
{\cal A}(s,t)=\int_{Q_0}^{\sqrt{s}}dk_1 {V(k_1)\over k_1^4}
\int_{Q_0}^{\sqrt{s}}dk_2 {V(k_2)\over k_2^4} T(k_1,k_2;s)
\end{equation}
$k_1$ and $k_2$ are the momenta entering the gluon ladder from either hadron,
$\sqrt{s}$ is the total energy, the two form factors
$V(k_i)$, $i$=1,2, represent the coupling of the proton to
the perturbative ladder via a
non-perturbative exchange, and the $1/k_i^4$ come from the propagators of the
external legs. $T(k_1,k_2;s)$ will obey the BFKL equation both for $k_1$ and
$k_2$, and the two independent evolutions will be related by the
driving term $T_0$ representing the 2-gluon exchange
contribution and thus proportional to $\delta(k1-k2) s^{1+\epsilon_0}$.
The next terms $T_n$ will be given by Equation (2) but cut off at small
$k$:
\begin{equation}
T_n(k_T,k_2;s)=\theta(\sqrt{s}>k_T>Q_0) \hat K T_{n-1}(k_T',k_2;s')
\end{equation}
Under these assumptions, and working at fixed $\alpha_s$, one can show that
the amplitude (6) conserves the structure found in \cite{CollinsL}:
\begin{equation}
{{\cal A}\over s}=C_0 s^{\epsilon_0}+\sum_{n=1}^\infty C_n(s) s^{\epsilon_n(s)}
\end{equation}
This solution reduces to the usual solution of the BFKL equation when
$s\rightarrow\infty$ and $Q_0\rightarrow 0$. The coefficients $C_n$ depend on
the model assumed for the coupling $V(k)$ between the non-perturbative and the
perturbative physics and their $s$ dependence is a threshold effect coming
from the integration in (6). Their only general property is that they are
positive.
On the other hand, the powers $\epsilon_n(s)$ are universal functions that
depend only on $\alpha_S$ and $\sqrt{s}/Q_0$.

\section{Interplay between soft and hard QCD: a model}
As the coefficients of the series (8) are model-dependent, we do not attempt to
calculate them, but rather try to assess the constraints that present
data place on them. We shall then be able to decide whether such
perturbative effects could play a substantial role
in soft physics at future colliders.
As all the $C_n$ are positive, the behaviour of the series (8) will not be
very different from that of its leading term, and so we truncate it. We
also make an educated guess for the threshold
function contained in $C_1(s)$. This does not affect our results for the
values of $Q_0$ shown here. We finally impose crossing symmetry to get
the real part of the amplitude. This gives
\begin{eqnarray}
{\tilde{\cal A}\over s}&=&C_0s^{\epsilon_0} +[c_1 (1-{Q_0\over
\sqrt{s}})^2\ \theta(\sqrt{s}-Q_0)]\ s^{\epsilon_1(s)}\\
{\cal A}(s)&=&{\tilde{\cal A}(s)}+\tilde{\cal A}(s e^{-i\pi})
\end{eqnarray}
with $c_1$ a positive constant.
To calculate $\epsilon_1(s)$ we assume that $Q_0$ is
the scale of $\alpha_S$ and
take $\Lambda_{QCD}=200$ MeV. Using the results of reference \cite{CollinsL},
we calculate the curves of Figure 3, for various values of the cutoff $Q_0$
and thus of $\alpha_S$. One sees that the effective power is much smaller than
its purely perturbative counterpart (4), {\it e.g.} for $Q_0$=2 GeV, the usual
estimate (4) gives $\epsilon$=0.8, whereas a cut-off equation gives values half
as big at accessible energies.

Again, we consider both a pole fit and an eikonalized one.
Note that the use of such an
eikonal formalism \cite{MargolisF} is not derived from QCD. In
fact, the BFKL equation in principle sums multi-gluon ladders in the $s$
and $t$ channels, so that in the purely perturbative case the eikonal
formalism is probably too na\"\i ve. However, in this case, it can be thought
of as an expansion in the number of form factors $V(k_1) V(k_2)$. This is
definitely not included in the BFKL equation. We further
add the meson trajectories of (1) to the amplitude (9),
and proceed to fit the data.

The first obvious observation is that the extra perturbative terms do not help
the fit: due to the positivity of the $C_n$, they cannot produce a bump in
$\rho$ that would explain the UA4 measurement.
So, one gets the best fit when the new QCD terms are actually
turned off. We want to examine here what constraints are placed
on them by present
data, and so we proceed as follows: we choose two values of $Q_0$, 2 and 10
GeV, where one could imagine to cut-off perturbative QCD. We then proceed to
fit the data, for increasing values of $c_1$, letting all other parameters
free. When we reach the 90\% confidence letter (C.L.)
as defined by our $\chi^2$/d.o.f., we have
found the highest perturbative contribution permissible. For $Q_0=2$ GeV, this
gives $c_1/C_0\leq 6\times 10^{-4}$ for the pole fit, and
$2\times 10^{-3}$ for the
eikonal one. For $Q_0=10$ GeV, the corresponding
values are $8\times 10^{-3}$ and
$11\times 10^{-3}$. We plot the resulting curves in Figures 1 and 2.

The parameters of the non-perturbative component are modified from
those of Table 1 by a few percent only. The coupling $C_0$ is increased a
little while the power $\epsilon_0$ goes down. This maximizes the perturbative
contribution, which is mainly constrained by the Tevatron measurement of the
total cross section. One sees that the small perturbative coupling can lead to
quite dramatic consequences at the SSC and LHC. We show in Table 2 the 90\%
C.L. on the total cross section and the $\rho$ parameter at future colliders.
As the perturbative contribution can become very large, the effect of
unitarization is non negligible, and depending on the value of $Q_0$, the
total cross section
could reach values as high as 230 mb at the SSC, with half of it coming from
perturbative resummation.

\begin{center}
\begin{tabular} {|l|l|l|l|l|}   \hline
collider/& pole fit     & eikonal fit & pole fit    & eikonal fit \\
quantity & $Q_0=2$ GeV   &$Q_0=2$ GeV&$Q_0=10$ GeV & $Q_0=10$ GeV     \\ \hline
SSC $\sigma_{tot}$(mb) &565 (459)& 228 (133)&180 (71)     & 139 (34)\\ \hline
LHC $\sigma_{tot}$(mb) &231 (138)  & 160 (74)&130 (35)     &113 (19) \\ \hline
Tevatron $\sigma_{tot}$(mb) &77 (9)   & 77 (10)&75  (6)      &74 (5) \\ \hline
SSC $\rho$&0.90  & 0.23 &0.27         &0.26 \\ \hline
LHC $\rho$&0.66  & 0.24 &0.21         &0.20 \\ \hline
Tevatron $\rho$  &0.20  & 0.16        &0.14         &0.12 \\ \hline
\end{tabular}
\end{center}

\noindent Table 2: Allowed values of the cross section and the $\rho$
parameter. The first two columns correspond to an infrared cutoff $Q_0=2$ GeV
and the two last ones to $Q_0=10$ GeV, see Equation (9). The number in
parenthesis next to the cross section is the value of the perturbative
component.

We emphasize that the estimates of Table 1 are conservative, as they
correspond to cutoffs of 2 and 10 GeV. Cutting off the
evolution when $\alpha_S\approx 1$ would give a total cross section of at least
3 b at the SSC, and be consistent with all available collider and fixed target
data!

\section{Conclusion}
We have shown that the BFKL equation can be used to evolve the soft pomeron to
higher $s$, and that perturbative effects could become measurable at the
SSC/LHC. These effects are cutoff dependent, and perturbative physics seems to
couple very weakly to the proton in the diffractive region, its coupling
strength being a few percent of that of the soft pomeron. However,
even a very weak coupling turning on at an energy
of a few GeV can lead to measurable
effects at sufficiently large energy. It is known that the pomeron couples to
quarks, and quarks to gluons. Therefore, the coupling to the BFKL ladder
cannot be zero, and specific models can be built for it \cite{cudell}.

This contribution is genuinely new and comes entirely from a QCD analysis. One
should not be misled by previous parton models \cite{Margolis} which, while
using a partonic picture, keep it mostly non-perturbative, replacing the small
power $s^{\epsilon_0}$ of (1) by a small power $x^{-\epsilon_0}$ in the gluon
structure function $xg(x)$. In the present model,
$xg(x)$ will contain the same powers
$\epsilon_i(Q_0/x)$ as the total cross section,
but their coefficients will in general be different from those entering
the total cross section, and the relation between them will be model dependent.

The existence of such possibilities, and the fact that very large total cross
sections are expected from the same kind of arguments that
lead one to predict a
rising cross section \cite{MargolisF,ChengW}, shows that small momentum physics
contains a wealth of open possibilities worth exploring experimentally. It
also suggests that a large proportion of events could become calculable at very
high energy, and so could be used for the detection of new physics.

\section*{Acknowledgments}
This work was supported in part by NSERC (Canada) and les fonds FCAR
(Qu\'ebec).

\newpage
\section*{Figure Captions}

\noindent Figure 1: Pole fits (1) and (10) at zero momentum
transfer, for pp (squares) and $\rm{\bar p}$p (crosses)
scattering. The lowest curve
is the best (non-perturbative) fit, while the two upper
curves are allowed by a cut-off BFKL
equation at the 90\% C.L., for an infrared cutoff $Q_0$=2 GeV or
10 GeV, as indicated. (a) shows the total
cross section and (b) the ratio of the real-to-imaginary parts of the
amplitude. The data are from reference \cite{data}.

\hfill\break
\noindent Figure 2: Same as Figure 1, but after eikonalization.

\hfill\break
\noindent Figure 3:  The effective power of s of Equation (8) that
results from a cut-off BFKL equation, for various values of the infrared cutoff
$Q_0$, as indicated next to the curves.

\end{document}